\documentstyle[12pt,cite,epsf,epsfig]{article}
\newcommand{\newc}{\newcommand}
\newc{\sgn}{\mr{sgn}\,}
\newc{\ra}{\rightarrow}
\newc{\rpv}{$\mathrm{\not\!R_p}$}
\newc{\met}{$\not\!\!E_T$}
\newc{\rp}{$\mathrm{R_p}$}
\newc{\real}{\mathcal{R}e}
\newc{\alsm}{{\displaystyle \sum_{\alpha=1,2}}}
\newc{\besm}{{\displaystyle \sum_{\beta=1,2}}}
\newc{\al}{\alpha}
\newc{\ga}{\gamma}
\newc{\de}{\delta}
\newc{\cw}{\cos\theta_w}
\newc{\ssw}{\sin^2\theta_w}
\newc{\ccw}{\cos^2\theta_w}
\newc{\cbe}{\cos\beta}
\newc{\sbe}{\sin\beta}
\newc{\sh}{\hat{s}}
\newc{\sa}{\sin\al}
\newc{\ca}{\cos\al}
\newc{\bv}{$\mathrm{\not\!B}$}
\newc{\lv}{$\mathrm{\not\!L}$}
\newc{\ie}{{\it i.e.\/}\ }
\newc{\lam}{\lambda}
\newc{\cht}{\tilde{\chi}}
\newc{\upt}{\tilde{u}}
\newc{\elt}{\tilde{\ell}}
\newc{\hgt}{\tilde{H}}
\newc{\nut}{\tilde{\nu}}
\newc{\dnt}{\tilde{d}}
\newc{\psb}{\bar{\psi}}
\newc{\rtt}{\sqrt{2}}
\newc{\mut}{\tilde{\mu}}
\newc{\mr}{\mathrm}
\newc{\bath}{\bar{\theta}}
\newc{\tht}{\theta}
\newc{\JC}{{\bf J}}
\newc{\lra}{\longrightarrow}
\newc{\eg}{{\it e.g.\,}}
\newc{\barr}{\begin{eqnarray}}
\newc{\earr}{\end{eqnarray}}
\newc{\beq}{\begin{equation}}
\newc{\eeq}{\end{equation}}
\newc{\me}{\mathcal{M}}
\newc{\dbm}{\partial_\mu}
\newc{\sgm}{\sigma_\mu}

\newcommand{\Slash}{{\slash \!\!\!}}
\newcommand{\lnr}{{\ln\left({q^2 \over  \mu^2}\right)}}
\newcommand{\lnrr}{{\ln^2\left({q^2 \over  \mu^2}\right)}}
\def\msnu{m_{\tilde \nu}}
\def\barr{\begin{array}}
\def\earr{\end{array}}
\def\be{\begin{equation}}
\def\ee{\end{equation}}
\def\ra{\rightarrow}
\def\dis{\displaystyle}
\catcode`@=11 
\def \gsim{\mathrel{\mathpalette\@versim>}}
\def \lsim{\mathrel{\mathpalette\@versim<}}
\def \@versim#1#2{\lower0.4ex\vbox{\baselineskip\z@skip\lineskip\z@skip
     \lineskiplimit\z@\ialign{$\m@th#1\hfil##\hfil$%
     \crcr#2\crcr\sim\crcr}}}
\catcode`@=12 
\def\gev{\: \rm GeV}
\def\tev{\: \rm TeV}

\def\etal{{\em et al.}}

\begin{document}
\setcounter{page}{0}
\renewcommand{\thefootnote}{\fnsymbol{footnote}}
\thispagestyle{empty}

\begin{titlepage}
\vspace{-2cm}
\begin{flushright}
MRI--P--020704\\[2ex]
{\large \tt hep-ph/0207247}\\
\end{flushright}
\vspace{+2cm}

\begin{center}
 {\Large{\bf QCD Corrections to Resonant Slepton Production in 
     Hadron Colliders}}\\
\vskip 0.6 cm
{\bf Debajyoti Choudhury, Swapan Majhi {\rm and} V. Ravindran}
        \vskip 1cm
{\it Harish-Chandra Research Institute, Chhatnag Road, Jhusi, Allahabad 211 019,
        India}
        \vskip 0.5cm
{E-mail: debchou,swapan,ravindra@mri.ernet.in}
\\
\vskip 0.2 cm
\end{center}
\setcounter{footnote}{0}
\begin{abstract}\noindent
We consider resonant production of sneutrino and slepton at hadronic colliders
such as the Tevatron and the LHC within the context of a $R$-parity violating
supersymmetric model.  We present next to leading order QCD corrections
to total cross sections which originate from both quark- as well as
gluon-initiated processes.  For couplings involving only the first generation 
quarks, the $K$ factor at the Tevatron can be as large as 1.5 for a 
$100 \gev$ sfermion and falls to nearly 1.1 as the sfermion mass reaches 
$1 \tev$. At the LHC, the variation is between 1.2 and 1.45 for masses less 
than 2 TeV. While the dependence on the parton density parametrization is 
found to be mild, this ceases to be true if the strange quark plays a 
dominant role in the production process. We also study the renormalization 
and factorization-scale dependence and find it to be less pronounced for 
the NLO cros sections as compared to the LO. The results obtained in this 
article are also applicable to resonant production of any color-neutral 
scalar. 
\end{abstract}
\end{titlepage}

\setcounter{footnote}{0}
\renewcommand{\thefootnote}{\arabic{footnote}}

\setcounter{page}{1}
\pagestyle{plain}
\advance \parskip by 10pt
\section{Introduction}

The scalar sector of the Standard Model
(SM), while being integral to the validity of this otherwise eminently
successful model, has also been somewhat of an embarrassment.
Quite apart from the fact that the Higgs particle has, till date, defied
all attempts at detecting it, there is the theoretical problem that the
mass of this particle is not protected by any symmetry (at least, not
within the SM). Consequently, quantum
corrections would tend to drive it to the next higher scale of
interaction, an eventuality that, apart from running counter to the
indications coming from electroweak precision
measurements\cite{EWWG}, would also
lead to a loss of perturbative unitarity.  To overcome this
as well as certain other lacunae of the SM, many
models going beyond the SM have been proposed.
Two of the most attractive classes of such models
comprise those incorporating supersymmetry~\cite{SUSY} and/or
grand unification~\cite{GUT} (especially scenarios with a low
intermediate scale~\cite{framp}).  Such models, however, predict,
additional particle states, including scalars.  What is
most interesting is that the coupling of the first
generation SM fermions to these scalars need no longer be suppressed, thus
offering hope for novel signatures.

The last-mentioned feature has, naturally, attracted much attention,
especially in the context of the current and future colliders.
Apart from pair-production (determined, in the most part, by the
gauge interactions), an enhanced coupling to fermions opens up the
possibility of resonance production at colliders whether
hadronic
($p \bar p$ or $p p $)
\cite{slepton_squark_reso,slepton_reso,squark_reso},
$e^+ e^-$~\cite{resonance_ee,Barger:1989rk}
or $e^\pm p$~\cite{HERA_expl}.
These studies conclude that not only is discovery
guaranteed for a significantly large part of the parameter
space, even a measurement of the coupling strengths to a reasonable degree
of accuracy might be possible.

Most of these analyses, however, have been performed only at the
Born level. In view of the interesting
consequences, it is desirable that quantum corrections
to such resonance production processes should be investigated.
While this has been done
in the context of $e p$ colliders~\cite{NLO:ep},
a similar exercise has not been
attempted for hadronic colliders. In this paper, we seek to rectify
this lacuna.  Before we embark on
such a venture, however, it is important to note that, in a generic
model going beyond the SM, scalars could appear in many a  hue. The
quantum numbers are of crucial importance as these would determine not only
the production cross-sections but also the dominant decay channels and
hence the possible modes of discovery.
For the sake of concreteness, we shall confine ourselves to
a discussion of sneutrino and slepton production in the context of
a $R$-parity violating supersymmetric model.
While, at first, this may seem to be a very restrictive assumption,
in reality, these constitute
very typical examples of {\em colour-neutral scalars}.
One might as well have considered a generic multi-higgs model
wherein some of the low-lying states have an enhanced coupling
with the lighter fermions.
The prime rationale behind our choice is that
while the scalar masses can be protected naturally in supersymmetric
models,  the same is not so straightforward in non-supersymmetric
models (grand unified or otherwise).
Moreover, the $R$-parity violating Minimal Supersymmetric
Standard Model (MSSM) being a richer (low-energy) theory,
offers a larger set of possibilities,
both in the context of the neutrino anomalies
seen at {\sc kamiokande}\cite{kamiokande} or {\sc karmen}\cite{karmen}
or the unexplained high-$Q^2$ events at
{\sc hera}\cite{HERA_expl}.

The plan of the paper is as follows.
We start this article (Section~\ref{sec:rpv})
with a brief review of the status of
$R$-parity conservation within the MSSM..
Section~\ref{sec:LO} describes the particular
resonance production processes (at the lowest-order) that we are
interested in. The formalism and the calculations for the NLO
corrections are set out in the following section.
In section~\ref{sec:results}, we present the numerical results
and a discussion thereof. And finally we summarise.

\section{$R$-parity violation: a mini-review}
        \label{sec:rpv}

As is well known, within the SM, both
baryon ($B$) and lepton ($L$) number conservation
are but accidental consequences of the choice of the particle
content\footnote{Indeed, non-perturbative effects within the SM itself
        do break $B + L$ symmetry.}.
In extensions of the SM, such an accidental occurrence is obviously not
guaranteed. For example, in a generic grand unified theory (GUT),
both the gauge and the scalar sector interactions
violate each of $B$ and $L$.
This is potentially catastrophic as a simultaneous breaking of
both $B$ and $L$ could lead to rapid proton decay.
Within GUTs, however, gauge boson-mediated proton decay is
naturally suppressed on account of the large gauge-boson masses.
On the other hand, the scalar sector
has to be carefully chosen so as to suppress any effective operator
leading to proton decay.

Within the context of the MSSM though, we do not have
the option of demanding the sfermion or gaugino fields
to be superheavy.
However, a similar suppression can be achieved
by introducing a discrete symmetry, $R \equiv (-1)^{3 (B - L) + 2 S}$
(with $S$ denoting the spin of the field)~\cite{fayet} that
serves to rule out both $B$ and $L$ violating terms in the superpotential.
In addition, this symmetry renders the lightest
supersymmetric partner absolutely stable. The introduction of
this symmetry is clearly an
{\em ad hoc} measure and is not even strictly essential to rule out
proton decay. Hence, it is of interest to consider possible violations
of this symmetry especially since it has rather important experimental
consequences, not the least of which concerns the detection of the
supersymmetric partners.

The possible $R$-parity violating (\rpv)
terms in the superpotential can be parametrised as
\be
    {\cal W}_{\not R_p} = \mu_i L_i H_2
                        + \lambda_{ijk} L_i L_j E^c_k
                        +  \lambda'_{ijk} L_i Q_j D^c_k
                        +  \lambda''_{ijk} U^c_i D^c_j D^c_k \ ,
      \label{eq:superpot}
\ee
where $L_i$ and $Q_i $ are the $SU(2)$-doublet lepton and quark
superfields, $E^c_i, U^c_i, D^c_i$
the singlet superfields and $H_i$ the Higgs superfields.
Clearly $\lambda_{ijk}$ is antisymmetric
under the interchange of the first two indices, while $\lambda''_{ijk}$
is antisymmetric under the interchange of the last two.
Whereas the first three terms in eqn.(\ref{eq:superpot}) violate
$L$, the last term falls foul of $B$ conservation.
To circumvent the constraints imposed by the
non--observance of proton decay, we, thus, need to have
at least one of the two sets of couplings to be vanishingly small.
For the purpose of this paper, we assume that $B$ is a good symmetry
of the theory, or in other words
all of $\lambda''_{ijk}$ are zero.
This has the added advantage that all dimension six operators leading
to proton decay are suppressed~\cite{Ibanez:1992pr} along with the
dimension five ones. Such
a scenario might be motivated within certain theoretical
frameworks~\cite{hall-suzuki,Ibanez:1992pr} and also renders simpler the
problem of preservation of GUT--scale baryon asymmetry~\cite{baryo}.
Although the presence of the other \rpv\  terms could, in principle,
affect the baryon asymmetry of the universe,
such bounds are highly model-dependent and can
be evaded~\cite{dreiner-ross}. For example,
in cases where at least one $L$-violating
coupling involving a particular lepton
family is small enough ($ \lsim 10^{-7}$)
so as to (almost) conserve the corresponding lepton flavour
over cosmological time scales, such bounds are no longer effective.

Each of the terms in eqn.(\ref{eq:superpot}) has its unique set of
consequences, whether in low-energy phenomenology or in resonance
production. For example, while $\lambda_{ijk}$ lead to resonant
sneutrino production in
$e^+ e^-$ collider~\cite{resonance_ee,Barger:1989rk},
a non-zero $\lambda''_{ijk}$ leads to
resonant squark production in hadron--hadron collisions
  \cite{slepton_squark_reso,squark_reso}.
Even richer phenomenology is associated with the $\lambda'_{ijk}$ terms.
For example, the exchange of a $t$-channel sfermion alters
significantly~\cite{virtual}
both the rates and the kinematic distributions
of processes such as $\bar t t$ production or Drell-Yan production 
of dileptons.
More strikingly, a non-zero $\lambda'_{ijk}$ can lead to both
resonant squark production at an $e^\pm p$ facility~\cite{HERA_expl}
as well as to resonant slepton and sneutrino production at
a hadronic collider~\cite{slepton_squark_reso,slepton_reso}.
It is this last aspect that we shall concentrate on.

The relevant part of the Lagrangian can be written
in terms of the component fields as
\be
\barr{rcl}
   {\cal L}_{\lambda'} & = & \lambda'_{ijk}
            \left[ \overline{d_{kR} } \nu_{iL} \tilde{d}_{jL} +
                   \overline{d_{kR} } d_{jL} \tilde{\nu}_{iL} +
                   \overline{(\nu_{iL})^c } d_{jL} \tilde{d}^\ast_{kR}
            \right. \\[1.5ex]
& & \left. \hspace*{1.5em}
                   - \overline{d_{kR} } \ell_{iL} \tilde{u}_{jL} -
                   \overline{d_{kR} } u_{jL} \tilde{\ell}_{iL} -
                   \overline{(\ell_{iL})^c } u_{jL} \tilde{d}^\ast_{kR}
            \right] + {\rm h.c.}
\earr
      \label{lambda-pr}
\ee
Thus, while the squarks behave as leptoquarks in a
non-supersymmetric theory, the sleptons/sneutrinos behave as
if they are charged/neutral Higgses in a multi-Higgs-doublet
scenario. Clearly, non-zero values for these couplings could lead
to rather striking phenomenological consequences.
For example, pair production of squarks that subsequently decay through
an $L$ violating interaction, leads to a final state comprising a dilepton
pair along with jets~\cite{d0_rpar_lepq}. More interestingly,
the gluino production cross-section is larger and, in addition,
can lead to like-sign dileptons, thereby making the signal stand out even
more~\cite{Tev_likesign}. Non-observation of such signals thus rules out
a relatively light squark or gluino along with a sizable $\lambda'$.
However, these analyses can say very little about sleptons/sneutrinos
as the corresponding production cross-sections are much smaller
than those for a squark/gluino.
At an $e^+ e^-$ collider though, both pair production
of sleptons/sneutrinos and the corresponding backgrounds are weak processes
and hence such colliders are expected to be better suited for this
particular quest. Unfortunately, an $e^+ e^-$ collider energetic enough to
pair-produce sleptons is still very much in the future.

We now turn to the constraints from low-energy phenomenology.
Non-zero $\lambda'$s can lead, for example, to additional
four-fermi operators that may contribute to meson decays, neutral
meson mixings, some of which may be forbidden otherwise. Since
the absence of tree-level flavour changing neutral current processes
lead to rather severe constraints on the simultaneous presence
of more than one $\lambda'$~\cite{fcnc}, we shall henceforth
restrict ourselves to only one non-zero $\lambda'$. In
Table~\ref{tab:coup_limits}, we list the currently known bounds
on several of these couplings\footnote{A more complete list can be found
        in refs.\protect\cite{bounds}.}.
The strongest bound is on $\lambda'_{111}$
and is derived from non-observation of neutrinoless
double beta decay $(a)$~\cite{bb0nu}. The others are much weaker and
are derived from
        ($b$) upper bound on the mass of the
                $\nu_e$~\cite{hall-suzuki,belesev,gb_dc,anjan};
        data on ($c$) charged-current universality~\cite{Barger:1989rk};
                ($d$) atomic parity violation~\cite{APV};
                ($e$) $\tau \ra \pi \nu_\tau$ and
                      $D \ra K l \nu$~\cite{gb_dc};
        and     ($f$) $D^0$-$\overline{D^0}$ mixing~\cite{fcnc}.
\begin{table}[htb]
\begin{center}
\bigskip
\begin{tabular}{||c|c|| c|c|| c|c||}
\hline
$\{ijk\}$ & Existing bounds & 
$\{ijk\}$ & Existing bounds & 
$\{ijk\}$ & Existing bounds \\[1ex]
\hline
111 & 0.001$^{\:a)}$ &
     211 & 0.09$^{\:c)}$ &
        311  & 0.10$^{\:e)}$ \\
112 & 0.02$^{\:c)}$ &
     212 & 0.09$^{\:c)}$ &
        312  & 0.10$^{\:e)}$ \\
121 & 0.035$^{\:d)}$ &
     221 & 0.18$^{\:e)}$ &
        321  & 0.20$^{\:f)}$ \\
122 & 0.02$^{\:b)}$ &
     222 & 0.18$^{\:e)}$ &
        322  & 0.20$^{\:f)}$ \\
\hline
\end{tabular}
\caption[] {The  upper bounds on the $\lambda'$--type \rpv\ couplings
of interest for a common sfermion mass $\tilde{m} = 100$ GeV. 
The superscripts refer to the specific experiments 
leading to the constraints and as described in the text.
}
\label{tab:coup_limits}
\end{center}
\end{table}

Since these bounds are derived from effective 4-fermi operators,
they typically scale like the mass of the exchanged
sfermion\footnote{Of those listed in Table~\protect\ref{tab:coup_limits},
        the only exceptions to this rule are the bounds for
        $\lambda'_{111}$ and
        $\lambda'_{122}$~\protect\cite{bb0nu,hall-suzuki,belesev,gb_dc}.}.
Two points may be noted here. First, many of these bounds are actually
applicable only to particular combinations of couplings and masses and
reduce to those in the table only under the assumption of only one coupling
being non-zero. And secondly, in meson decays, most often it is the squark
that is exchanged; hence sleptons/sneutrinos could very well be much lighter
without contradicting the bounds.

\section{Leading order cross-section}
        \label{sec:LO}

The \rpv\ interaction Lagrangian, as presented in eqn.(\ref{lambda-pr}),
allows for the following resonance production processes at a hadronic
collider:
\be
\barr{rclcl}
        \lambda'_{ijk} & : & d_j + \bar d_k \ra \tilde \nu_i,
                       & \quad & d_k + \bar d_j \ra \tilde \nu^\ast_i \\
                       & : & d_k + \bar u_j \ra \tilde \ell_i,
                       &  & u_j + \bar d_k \ra \tilde \ell^\ast_i \\
\earr \ .
        \label{lambdaprime:processes}
\ee
The conjugate processes obviously have identical cross-sections at the
Tevatron, though not at the LHC.

Before we start, we will make a few simplifying assumptions.
Since QCD is flavour-blind, the {\em form} of the strong interaction
corrections would be independent of the particular initial state
quark in eqn.(\ref{lambdaprime:processes}). Hence, for the sake
of simplicity, we choose to
develop the formalism for the case of identical quarks.
Furthermore, rather than
consider a chiral coupling to the scalars, we assume that the
interaction is purely a scalar one. The chirality structure can
be accounted for at a later stage simply by introducing an extra
factor of 1/2. Note that neither of these
assumptions imply a loss of generality.

The leading order process of interest is then
\be
\bar{q}(p^{\prime})~ +~ q(p)~\rightarrow~ S(q)
        \label{Born_process}
\ee
where $p$ and $p^{\prime}$ are the momenta of incoming quark and
anti-quark respectively and $q$ that of the outgoing scalar.
The amplitude for this process is given by
\be
M^{(0)} = -i \lambda~\bar{v}(p^{\prime})~u(p)
\ee
where $\lambda$ is the scalar coupling constant.
Since we consider only light quarks in the initial state,
we have
\be
(p^{\prime})^2 = p^2 = 0
\ee
whereas the scalar has a (large) mass $m_q$.
The cross-section for the process in (eqn.\ref{Born_process})
is
\be
\sigma_{0}= {1 \over 3}~\frac{1}{4\,(p.p^{\prime})}\, \int\,
            \frac{d^{n-1}q}{(2\pi)^{n-1}\,2\,q_{0}}~
            {1 \over 4}|M^{(0)}|^{2}\,
            (2\,\pi)^n~\delta^n(p^{\prime} + p - q)
     \label{sig_Born}
\ee
where the factor $1/4$ arises from the spin averaging
for the incoming quarks and $4~p\cdot p'$ is the flux factor.
Taking the space-time dimension $n = 4$, the above reduces to
\be
  \sigma_{0} =   \lambda^{2} ~\frac{1}{3}~\frac{\pi}{2 s}~\delta(1-\tau)
\ee
where
\be
\tau = \frac{m^2_{q}}{s},
\hspace{1cm} s~ = (p^{\prime} ~+~p)^2,
\hspace{1cm} m_q^2 = q \cdot q \ .
\ee

\section{NLO corrections}
        \label{sec:NLO}
The QCD correction to the process of interest has contributions
from two different, but related, sources.
First, the quark-pair-initiated process
itself receives radiative correction.
To this must be added the
contribution arising from radiating off a soft gluon. And secondly, since
our true initial state is not quarks, but (anti-)protons, we must include
possible contributions from ``initial-state'' gluons as well. We consider
each in turn.

\subsection{Correction to the $q \bar q$ initiated process}
To calculate the QCD radiative correction
to this process, we start by computing the ${\cal O}(g_s^2)$
corrections to the vertex function and the self energy, where
$g_s$ is the QCD strong coupling constant.
A prime ingredient for this is the calculation of
the corresponding renormalisation constants
$Z_{\lambda}$ and $Z_2$. Even on regulating the ultraviolet (UV)
divergences, we would, expectedly, be left with
infra-red (IR) divergences, part of which will be cancelled once
we take into account the soft gluon bremsstrahlung contribution.
Throughout our calculation we shall use dimensional regularisation
to regulate any divergence and the $\overline {MS}$ prescription
for renormalising the results.

Let us first consider the vertex function $M^V$ upto order $g_s^2$.
This can be expanded as
\begin{equation}
M^V= M^{(0)} + M^{V(1)}
\end{equation}
where $M^{(0)}= -i \lambda~\bar{v}(p^{\prime})~u(p)$,  and
\begin{equation}
M^{V(1)} = -\lambda~ g_s^2 ~\frac{1}{(\mu^2)^{{n\over 2}-2}}~
\int  \frac{ d^nk}{(2\pi )^n}~ \frac{ \bar{v} (p^{\prime})~ t^a~t^a~
\Big[~\gamma ^{\mu} (-\Slash p^{\prime}  +  \Slash k   )
                    (\Slash p + \Slash k\, )  \,
                    \gamma _{\mu} ~ \Big]   \,u(p)}
      { [(p^{\prime} -k)^2  +  i  \eta~ ]\,
        [ (p +k)^2  +  i \eta~ ]\,
        [k^2  +  i  \eta~]  }
\ .
\end{equation}
The introduction of the (arbitrary) mass scale $\mu$
(called renormalisation scale)
is necessary to render
the strong coupling constant $g_s$ dimensionless in $n$ space-time
dimensions.  The
matrices $t^a$ are the Gell-Man matrices and satisfy
 $(t^a t^a)_{ij} = C_F \delta_{ij}$.

On using the equations of motion, the above can be simplified to
\begin{equation}
M^{V(1)}    =~ \bar{v}(p^{\prime})~(-i~\lambda)~\bar\Gamma^{(1)}~u(p) \ ,
\end{equation}
where
\begin{equation}
\bar \Gamma^{(1)}  =  -i g_s^2~C_F \frac{1}{(\mu^2)^{{n \over 2}-2}}~
 ~\Big[ -2  q^2~ I_1    +
        4 ~(  p -p^{\prime}  )^{\mu}~    I_{\mu}    +
        n ~  g^{\mu  \nu}~  I_{\mu \nu}  ~\Big ] \ ,
\end{equation}
with
\begin{eqnarray}
I_{\{1,~\mu,~\mu\nu \}}&=&\int  \frac{ d^nk}{(2\pi )^n}
\frac{\Big\{1,~k_\mu,~k_\mu~k_\nu\Big\}}{[(p^{\prime} -k)^2  +  i  \eta~ ]\,
[ (p +k)^2  +  i \eta ~]\,  [k^2  +  i  \eta~]  } \ .
\end{eqnarray}
Naive power counting shows that $I_{\mu \nu}$ is logarithmically
divergent in 4-dimensions, while the other two are convergent.
The integrals can be evaluated explicitly (for example,
using Feynman parametrisation) and the
results expressed in terms of Gamma functions.  The resultant
vertex function is then ($\epsilon \equiv n - 4$)
\begin{eqnarray}
\bar \Gamma^{(1)}&=&{\alpha_s \over 4 \pi} C_F\left({-q^2 \over 4 \pi \mu^2}
                       \right)^{{\epsilon \over 2}}
        { \Gamma^2\left(1+ {\epsilon \over 2} \right)
         \Gamma\left(1- {\epsilon \over 2} \right)
         \over \Gamma(2+\epsilon)}
\left [~-{2 \over \epsilon^2} ~(2+\epsilon)^2~\right] \ .
\end{eqnarray}

The renormalisation constant $Z_{\lambda}$ is defined through the relation
\begin{equation}
Z^{-1}_{{\lambda}} = 1 + \bar\Gamma^{(1)}|_{UV}
\end{equation}
and, of course, depends on the way the ultraviolet divergent part
$\Gamma^{(1)}|_{UV}$ is isolated.
Within the $\overline{MS}$ scheme, it can easily be ascertained to be
\begin{equation}
Z_{{\lambda}}  =  1 + \frac{\alpha_s}{4 \pi \Gamma(1+\epsilon/2)} C_F~\left({1 \over  4 \pi}
             \right )^{{\epsilon \over 2}}\left(~\frac{8}{\epsilon}\right)
          \ ,
        \label{Z_lambda}
\end{equation}
with
$\alpha_s \equiv g_s^2/4 \pi$.

The self energy correction to the Born amplitude (say, to the quark only)
can be expressed as
\begin{equation}
M^{S} =-i~\lambda
\bar v(p'){1 \over \Slash p}
\overline \Sigma(\Slash p)
u(p) \ ,
\end{equation}
where
\begin{equation}
\overline \Sigma(\Slash p)=
-i~g^2_{s}~ {C_F \over (\mu^{2})^{{n \over 2}  - 2}}
\int\,\frac{d^nk}{(2\,\pi)^n}
    \;
   \frac{\,\gamma^{\mu}~\left(\Slash p\, +\,\Slash k\right)~\gamma_{\mu}}
        {[k^2\,+\,i \eta\,]\,[\,(p+k)^2\,+\,i \eta\,]} \ .
\end{equation}
Notice that $\Sigma(\Slash p)$ does not contribute to the amplitude
given in eqn.(\ref{Born_process}) due to the massless nature of the light quarks.
On the other hand,  the above equation can be used to determine the
wave function renormalisation constant $Z_2$ through the relation
\begin{equation}
\frac{d\overline\Sigma(\Slash p)}{d\Slash p}|_{\Slash p=0}
        = i~(~Z^{-1}_{2} - 1) \ .
\end{equation}
In the $\overline{MS}$ scheme (and in a scale independent way)
this can be rewritten as
\begin{equation}
Z_2=1+ {\alpha_s \over 4 \pi \Gamma(1+\epsilon/2)} C_F
        \left({1 \over 4 \pi} \right)^{{\epsilon \over 2}}
\left( {2 \over \epsilon}  \right) \ .
        \label{Z_2}
\end{equation}

Our next task is to compute the virtual contributions to the process
given in eqn.(\ref{Born_process}).  In order to do this, we have to
redefine the fields
and the coupling constants in terms of the renormalised ones
(and, of course,  the renormalisation constants  $Z_2$ and $Z_{\lambda}$.)
This is equivalent to adding UV counter terms corresponding to
the vertex function and self energy contribution.
Note that self energy contribution to the amplitude is identically
zero due to the on-shell condition.   Hence only vertex function and the
counter terms contribute to the amplitude, and
\begin{equation}
M^{(0)+vir+CT}=M^{(0)}+M^{V(1)}+M^{CT} \ .
\end{equation}
It turns out that the effect of the counter term(CT) is
\begin{equation}
M^{(0)}+M^{CT}=(-i~ \lambda_R)~ {Z_{\lambda} \over Z_2}~ \bar v(p')~ u(p)
\end{equation}
where the renormalised coupling
$\lambda_R$ is related to the   unrenormalised one through
\begin{equation}
\lambda_{R} = \lambda~\frac{Z_{2}}{Z_{{\lambda}}}\ .
\end{equation}
The virtual and counter term contribution to the Born diagram can be
expressed as
\begin{eqnarray}
\sigma^{(0)+vir+CT}= {1 \over 3}~
        {1 \over 4 \big(p \cdot p'\big)} \int {d^{n-1} q \over (2 \pi)^{n-1} 2 q_0}
{1 \over 4}|M^{(0)+vir+CT}|^2
(2 \pi)^n \delta^n(p+p'-q) \ .
      \label{sig_0_vir_CT}
\end{eqnarray}
Substituting eqns.(\ref{Z_lambda},\ref{Z_2}) in eqn.(\ref{sig_0_vir_CT}), we have
\begin{equation}
\sigma^{(0)+vir+CT}= {1 \over 3}~
{\pi \over 2 s} \lambda_R^2 \Bigg(1+{\alpha_s \over 4 \pi} C_F F^{(1)}\Bigg)
\delta(1-\tau) \ ,
\end{equation}
where
\begin{equation}
F^{(1)}={4 \pi \over \alpha_s C_F} \Bigg( {Z_{\lambda}^2 \over Z_2^2}
           + 2 \bar \Gamma^{(1)} -1\Bigg) \ .
\end{equation}
Expanding around $\epsilon=0$ and
neglecting those terms which vanish in the limit $\epsilon \rightarrow 0$
we get
\begin{equation}
F^{(1)}={ 2 \over \Gamma\left (1 + {\epsilon \over 2} \right)}
~({1 \over 4 \pi})^{\epsilon \over 2}
\Bigg[ -{8 \over \epsilon^2} ~+~{6 \over \epsilon}~ -~{4 \over \epsilon}
~\lnr
-\lnrr-2 +\pi^2\Bigg] \ .
\end{equation}

Next, we compute the contribution from
the gluon bremsstrahlung to ${\cal O}(\alpha_s)$:
\begin{equation}
\bar q(p')~+~q(p) \rightarrow S(q)~+~g(k).
\end{equation}
Here, $k$ is the momentum of the out going gluon.
The corresponding cross-section is
\begin{equation}
\sigma_r={1 \over 3}~ {1 \over 4 \big(p \cdot p'\big)} \int~ dPS_2~{1 \over 4} |M_r|^2
\end{equation}
where the two body phase space is
\begin{equation}
dPS_2= \frac{d^{n-1}k}{(2 \,\pi)^{n-1}\,2\, k^0}~
              \frac{d^{n-1}q}{(2 \,\pi)^{n-1}\,2\, q^0}
                              ~(2\,\pi)^n~\delta^n(\,k + q - p - p'\,)
\end{equation}
and the amplitude $M_r$ is given by
\begin{equation}
M_r = -\,\lambda\,g_s\,\bar{v}(p^{\prime})~t^a
        \Big[~\frac{\gamma^{\mu}~(\Slash k- \Slash p^{\prime})\,}
         {(k-p^{\prime})^2}\,+\,\frac{(\Slash p- \Slash k)~
          \gamma^{\mu}}{(p - k)^2}\,~\Big]~u(p) ~\varepsilon_{\mu}^* (k) \ .
\end{equation}
The phase space integration can be easily done in the
centre of mass frame of incoming quarks wherein the
momenta of the particles can be parametrised as
\begin{equation}
\barr{rclcrcl}
p &=& \dis \frac{\sqrt s}{2}\,\Big(1,0,0,1\Big),  & \qquad&
 p^{\prime} &= & \dis \frac{\sqrt s}{2}\, (1,0,0,-1)
 \\[2ex]
 k &=& |q| \Big(1, \sin \theta, 0, \cos\theta\Big), & \qquad&
 q &=& \dis \Big (q_{0}, -|q| \sin \theta, 0, -|q| \cos\theta\Big)
  \\[2ex]
q_{0} & = & \dis {\sqrt{s} \over 2} (1+\tau)
   & & |q| & =& \dis {\sqrt{s} \over 2} (1-\tau),
\earr
\end{equation}
with the square of the matrix element being given by
\begin{equation}
|M_r|^2 = 4 \lambda^2 g_s^2 C_F
        \;
        \left(n-2-{4 \over 1-\tau} +{4 \over (1-\tau)^2} \right)
         \left ({1 \over 1+y} + { 1 \over 1-y} \right)  \ .
  \label{matrix_sq}
\end{equation}
Substituting the above in eqn.(\ref{matrix_sq})
and performing the integration, we have
\begin{eqnarray}
\sigma_{r} &=& {1 \over 3}~
{\alpha_s \over 4s} C_F \lambda^2
{(4 \pi)^{- \epsilon / 2} \over
\Gamma\left(1+{\epsilon \over 2}\right)}
\Bigg[
\Bigg({8 \over \epsilon^2} + {4 \over \epsilon}~ \lnr +\lnrr
-{\pi^2 \over 3}\Bigg)\delta(1-\tau)
\nonumber \\[2ex]
&&+{4 \over \epsilon} { 1+ \tau^2 \over (1-\tau)_+}
+2 (1-\tau)
+4 (1+\tau^2) \left( { \ln(1-\tau) \over (1-\tau)} \right)_+
\nonumber \\[2ex]
&&+2 {1+\tau^2 \over (1-\tau)_+} \lnr
-2 {1 + \tau^2 \over (1-\tau)} \ln(\tau) \Bigg]
   \ .
\end{eqnarray}
In the above, the $``+"$ prescription for a function $f(z)$ is defined as
\begin{equation}
\int_0^1 dz {f(z) \over (1-z)_+} = \int_0^1 dz {f(z)-f(1) \over (1-z)} \ .
\end{equation}
Notice that the above integral is divergent as $z \rightarrow 1$.
This is but a manifestation of the
collinear divergence arising due to the masslessness of the
incoming quarks,
and can be safely absorbed into unrenormalised parton densities by a suitable
counter term.  Within the $\overline {MS}$ scheme,
the said counter term is given by
\begin{eqnarray}
\sigma^{ct}&=&{1 \over 3}~{\alpha_s \over 4 s} C_F \lambda_R^2
     {1 \over \Gamma\left(1+{\epsilon \over 2}\right)}~
\left({M^2 \over 4 \pi \mu^{2}} \right)^{\epsilon \over 2}~
\Bigg [- 6 ~\delta(1-\tau)
- {4 \over \epsilon}~ {1+\tau^2 \over (1-\tau)_+}
\Bigg ] \ ,
        \label{sig_ct}
\end{eqnarray}
where $M$ is the factorisation scale.
Adding the virtual corrections to the bremsstrahlung contribution with the
collinear counter term (eqn.(\ref{sig_ct})), we get,
upto ${\cal O}(\alpha_s)$,
\begin{eqnarray}
\sigma^q&=&{1 \over 3}~ {\pi \over 2 s} \lambda_R^2 \delta(1-\tau)+
{1 \over 3}~
{\alpha_s \over 4s} C_F \lambda_R^2
\Bigg[
\Bigg( -3~ \lnr -2  +{2\pi^2 \over 3}\Bigg)\delta(1-\tau)
\nonumber \\[2ex]
&&
+3~\delta(1-\tau) \ln\left(q^2 \over M^2 \right)+2 (1-\tau)
\nonumber \\[2ex]
&&
+4 (1+\tau^2) \left( { \ln(1-\tau) \over (1-\tau)} \right)_+
+2 {1 + \tau^2 \over (1-\tau)_+} \ln\Big({q^2 \over M^2\,\tau}\Big) \Bigg]
   \ .
        \label{sig_nlo_q}
\end{eqnarray}

\subsection{Contribution from the gluon initiated process}
We now compute the final piece, namely the contribution of
the Compton-like process to order $\alpha_s$:
\begin{equation}
g(k)~+~q(p) \rightarrow S(q)~+~q(p').
\end{equation}
The cross-section is given by
\begin{equation}
\sigma_c= {1 \over 4 \big(p \cdot k\big)}
{1 \over 8}~{1 \over 3}
\int~ dPS_2~{1 \over 2 (n-2)} |M_c|^2
        \label{sig_glu_1}
\end{equation}
with the amplitude $M_c$ being
\begin{equation}
M_c = -\,\lambda\,g_s\,\bar{u}(p^{\prime})~t^a
        \Big[~\frac{\gamma^{\mu}~(\Slash p'- \Slash k)\,}
         {(p'-k)^2}\,+\,\frac{(\Slash p+ \Slash k)~
          \gamma^{\mu}}{(p + k)^2}\,~\Big]~u(p) ~\varepsilon_{\mu} (k)
        \ .
\end{equation}
Once again, the two-body phase space ($dPS_2$)
integration can be easily done in the
centre of mass frame of incoming gluon and quark wherein
\begin{equation}
|M_c|^2 = 4 \lambda^2 g_s^2
\Bigg[
\left(-4 \tau + { (n-2) \over 1-\tau} \right) {4 \over 1+y}
+ y~ (1-\tau) (n-2)
+6 +10 \tau -3 n -n \tau \Bigg] \ .
\end{equation}
On performing the angular integration, we get
\begin{eqnarray}
\sigma_c&=&{\lambda^2 g_s^2  \over 16 \pi s}~ {1 \over 3}~\left({q^2 \over 4 \pi \mu^2
         \tau}\right)^{{\epsilon \over 2}}
          {\Gamma\left(1+{\epsilon \over 2}\right)
             \over
          \Gamma\left(1 +\epsilon \right)
           }~
           {1 \over 4 \epsilon (2+\epsilon)}
\nonumber \\[2ex]
&& \times {1 \over (1-\tau)^{-\epsilon} }
\left[8 \Big(\tau^2+(1-\tau)^2\Big)
+4 \epsilon -{\epsilon~ (1-\tau) \over 1+\epsilon} \Bigg (
6 (1-\tau) +\epsilon (3+\tau) \Bigg) \right]
\nonumber \\[3ex]
 &=&
{\alpha_s \over 4s}~ {1 \over 3}~ \lambda^2 {1 \over
\Gamma\left(1+{\epsilon \over 2}\right)}~\left({1 \over 4\, \pi} \right)^{{\epsilon \over 2}}
\Bigg[
~{1 \over \epsilon}~\Big(\tau^2 +(1-\tau)^2\Big)
\nonumber \\[2ex]
&&+ {1 \over 2}~ \Big(\tau^2 +(1-\tau)^2\Big)
\ln\left({q^2 \over  \mu^2 }\right)~
 +{1 \over 2}~
\Big(\tau^2 +(1-\tau)^2\Big)
\ln\left({ (1-\tau)^2 \over \tau}\right)~
\nonumber \\[2ex]
&&+{1 \over 4}~(1-\tau)(7 \tau-3)~ \Bigg] \ .
\end{eqnarray}
With the $\overline {MS}$ counter term to remove the collinear divergence
coming from the massless incoming gluon and quark being given by
\begin{equation}
\sigma^{ct}={\alpha_s \over 4 s}~ {1 \over 3}~ \lambda_R^2
            {1 \over \Gamma\left(1+{\epsilon \over 2}\right)}
~\Bigg({M^2 \over \mu^2}\Bigg)^{\epsilon \over 2}
~\left( {1 \over 4 \pi}\right)^{\epsilon \over 2}~
\left [-{1 \over \epsilon}~\Bigg(\tau^2 +(1-\tau)^2\Bigg)
\right ] \ ,
        \label{sig_glu_ct}
\end{equation}
we finally have, to ${\cal O}(\alpha_s)$,
\begin{eqnarray}
\sigma^g&=& {\alpha_s \over 4s} ~{1 \over 3}~ \lambda_R^2
\Bigg[
{1 \over 2}~
\Big(\tau^2 +(1-\tau)^2\Big)
\ln\Bigg({q^2 \over  M^2 }\Bigg)~
\nonumber \\[2ex]
&&+{1 \over 2}~
\Big(\tau^2 +(1-\tau)^2\Big)
\ln\Bigg({(1-\tau)^2 \over  \tau}\Bigg)~
+ {1 \over 4}~(1-\tau)(7 \tau-3)~
\Bigg]
        \ .
        \label{sig_glue}
\end{eqnarray}
Note that both the quark and gluon initiated processes, after the
mass factorisation, are free of any IR divergences.
We use these results for our further analysis after folding
with appropriate parton distributions. For the numerical
calculation we made the renormalisation scale $\mu$ and the factorisation scale
$M$ equal (i.e. $\mu = M = \mu_{F}$).

\section{Results and Discussion}
        \label{sec:results}
Having obtained the analytic expressions in the last section, we now
endeavour to see the numerical size of these corrections. To this end,
one needs to define the
leading-order and the next-to-leading-order cross-sections
for a $p \bar p$ (or equivalently $p p$) collider:
\begin{tabular}{rp{5in}}
        $\sigma_{\rm LO}$ : & convolute the cross-section of
                                eqn.(\ref{sig_Born}) with
                             the appropriate LO quark distributions; \\
        $\sigma_{\rm NLO}$ : & convolute the cross-section of
                          eqns.(\ref{sig_nlo_q},\ref{sig_glue}) with
                             the appropriate NLO quark and gluon
                             distributions.
\end{tabular}

\noindent
Of course an additional factor of $1/2$ needs to be included in the
cross-sections to account for the chiral nature of the \rpv\
interactions.

Before we start in earnest, a minor digression.
Since QCD corrections are flavour-blind,
the {\em value} of the coupling $\lambda'$
is immaterial and only serves to set an overall
scale for the cross-section.
To be concrete, we shall choose to work with
\[
\lambda' = 0.01
\]
irrespective of flavour and the mass of the sneutrino/slepton.
While this may seem to be an inconsistent choice for a
light $\tilde e$ or $\tilde \nu_e$ (see Table~\ref{tab:coup_limits}),
this is not quite germane to the issue at hand.

\subsection{Sneutrino Production}
\begin{figure}[htb]
\centerline{\hspace*{3em}
\epsfxsize=17cm\epsfysize=8.0cm
                     \epsfbox{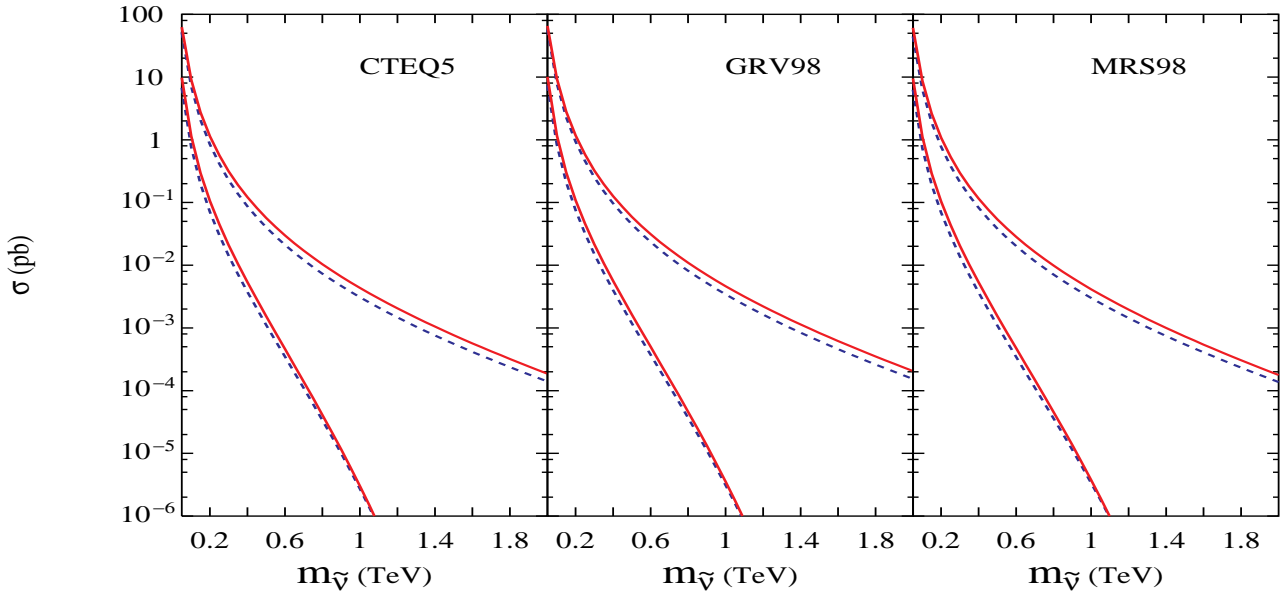}
}
\caption{\em Cross-section for resonant sneutrino production
             at the Tevatron (lower curves) and at the LHC (upper curves).
             The solid (dashed) curves represent the NLO (LO) cross-sections.
             The $R$-parity violating coupling $\lambda_{i 1 1}'$ has
             been set to 0.01. The three cases correspond to
             structure function parametrisations CTEQ5, GRV98 and
             MRS98 respectively.
 }
\label{fig:prod}
\end{figure}

To begin with, we concentrate on the resonance production of a sneutrino
starting with a $d \bar d$ initial state (at the Born level).
In Fig.\ref{fig:prod}, we plot both the
LO and the NLO cross-sections as a function of the sneutrino
mass and for three different choices of parton distributions.
The renormalisation scale $\mu_{F}$ as described
in the previous section has been chosen to be the same as the sneutrino mass.

%
On the face of it, the three sets of curves look quite similar,
a point that we shall remark on later. The dependence on the
scalar mass is
as expected and is occasioned by both the fall in the parton-level
cross-sections and the rapid decrease of the parton densities
at high momentum fractions. The latter effect, understandably, is
more pronounced at the Tevatron than at the LHC.

\begin{figure}[hbt]
\vspace*{-1ex}
\centerline{
\epsfxsize=14cm\epsfysize=10.0cm
                     \epsfbox{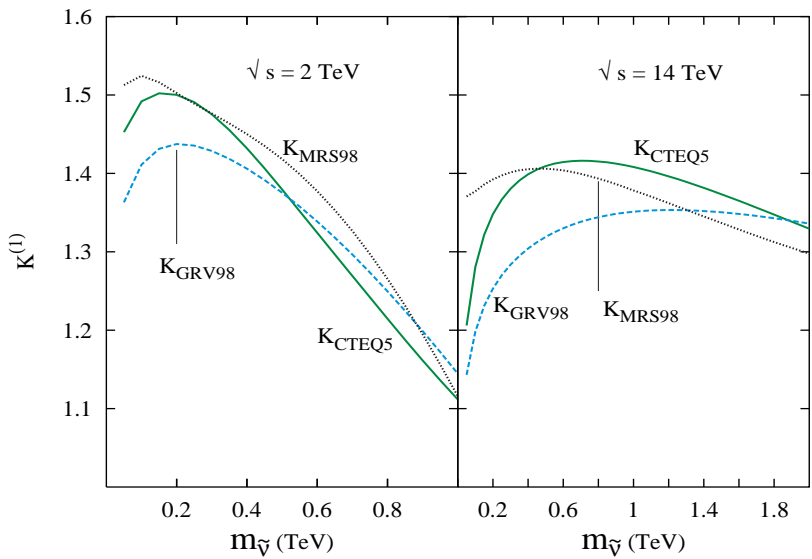}
}
\vspace*{-13 ex}
\caption{\em The $K$ factor for the process
             involving coupling $\lambda_{i 1 1}'$  as a function of
             the sneutrino mass as calculated for different parton
             distributions. The two graphs correspond to
             the Tevatron and the LHC respectively.
 }
\label{fig:K_fac}
\end{figure}

To parametrise the effect of the NLO corrections, it is common
to introduce the $K$-factor:
\be
        K \equiv \frac{\sigma_{\rm NLO}}{\sigma_{\rm LO}} \ ,
\ee
which we plot in Fig.\ref{fig:K_fac}. Let us concentrate
first on the results for the Tevatron. The near monotonic decrease
of $K$ with $\msnu$ is not unexpected. As $\msnu$ increases,
we are sampling increasingly larger values of parton momenta. This
has two immediate consequences. For one, the Compton contribution
becomes increasingly irrelevant. But more importantly, a large
$\msnu$ also means that the `primary quark' is less able
to radiate off a gluon. In fact, as $\msnu \ra \sqrt{s_{p \bar p}}/2$,
$K$ approaches unity.

It might seem significant that the $K$-factors, as calculated with
different sets of parton distributions, vary significantly amongst
themselves. This only reflects
the dependence of the cross-sections on our ans\"atz for the parton
densities. This assertion is supported by the fact the differences are
more pronounced for smaller sneutrino masses, where the cross-section
receives a larger contribution from relatively low momenta ($x_{\rm Bj}$)
partons and
hence prone to larger parametrisation errors.
Interestingly, the difference in the $K$-factors arise,
in a large measure, due to the differences in the LO cross-sections.
The relative differences between the cross-sections, as calculated
with the different Ans\"atze, actually decrease as we go from the
LO to the NLO calculations, and can be expected to become smaller
as progressively higher order corrections are incorporated.

Turning now to the results for the LHC, we see that,  for
$\msnu \gsim 300 \gev$, the behaviour is quite analogous to the
case of the Tevatron. Although the fall with the sneutrino mass
seems to be slower, it should be remembered that the graph
covers a much smaller range in $\msnu / \sqrt{s_{p p}}$.
The behaviour  at small masses ($\lsim 300 \gev$) seems puzzling
though. However, one must realize that the cross-section for
such light sneutrinos is dominated by low momenta
partons. And since existing data does not probe the parton densities
unto very low $x_{\rm Bj}$, the various Ans\"atze naturally have
differing predictions. Although it does not show up in the
curves of Fig.\ref{fig:prod}, again the difference in the
$K$-factor is dominated by the deviations in the LO cross-sections
rather than the NLO ones.


\begin{figure}[htb]
\vspace*{-7ex}
\centerline{
\epsfxsize=14cm\epsfysize=10.0cm
                     \epsfbox{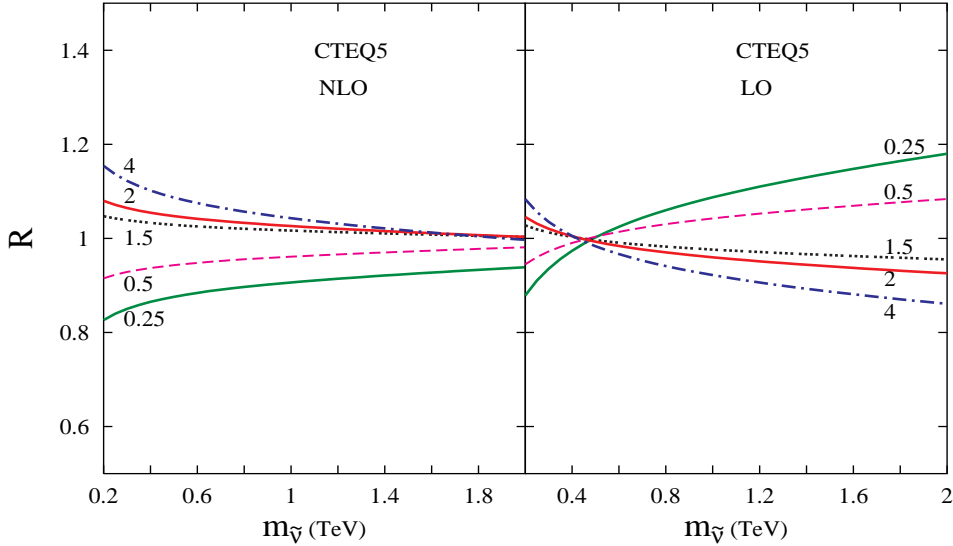}
}
\vspace*{-8ex}
\caption{\em The dependence of the cross-sections at the LHC
        on the value of the factorisation scale $\mu_{F}$. The ratio $R$
         (see eqn.(\protect\ref{the_ratio_R})) compares the cross
        section to the reference point of $\mu_{F} = m_{\tilde \nu}$.
        The legends on the graphs correspond to the ratio 
	$\mu_F / m_{\tilde \nu}$. 
	The left and right panels correspond to the NLO and LO
	cross-sections respectively. The CTEQ5 densities have been used.
 }
\label{fig:md_var}
\end{figure}


Having explored the dependence of the $K$-factor on the sneutrino
mass and the choice of parton densities, we now turn to the
final `unknown' viz. the renormalisation scale. Although the
most natural scale is that of the sneutrino mass (with many other similar
analyses making this choice too), the exact value of $\mu_{F}$ is somewhat
ambiguous. To quantify the ensuing dependence, we define the
ratio
\be \displaystyle
        R_{\rm NLO}(\mu_{F}; m_{\tilde \nu}) \equiv
                \frac{\sigma_{\rm NLO}(\mu_{F})}
                     {\sigma_{\rm NLO}(\mu_{F} = m_{\tilde \nu})}
        \label{the_ratio_R}
\ee
operative within a given parton density parametrisation. An analogous
expression can also be defined for the LO cross sections. 
In Fig.\ref{fig:md_var}, we exhibit the variation
of $R(\mu_{F}; m_{\tilde \nu})$ for both the LO and the NLO 
cross sections\footnote{While we show the dependence only for the CTEQ
 parametrization, those for the other parametrizations
 are very similar.}.
As the graph shows, the variation of the cross-section with
$\mu_{F}$ is relatively small. Furthermore, the variation 
reduces as one progresses from the LO calculation to the NLO.
The last observation lends hope that the remaining 
scale ambiguity can, presumably,
be reduced by adding still higher order corrections. 

\subsubsection{Initial states with strange quarks}
\begin{figure}[htb]
\centerline{\hspace*{3.0em}
\epsfxsize=17cm\epsfysize=8.0cm
                     \epsfbox{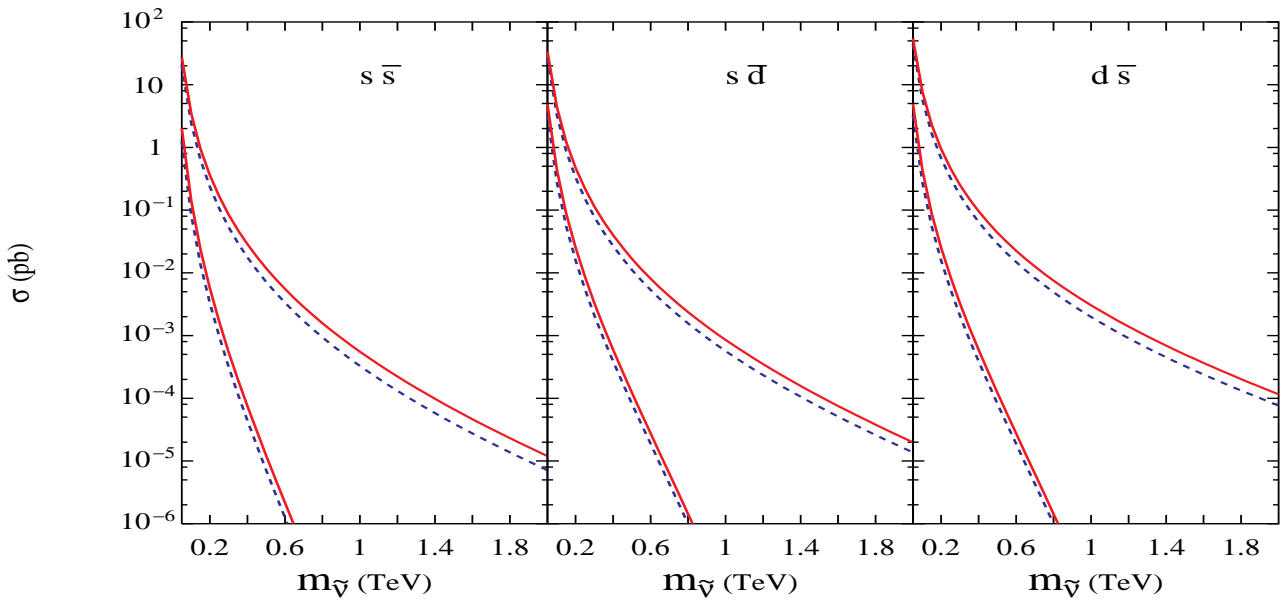}
}
\caption{\em Cross-section for resonant sneutrino production
             at the Tevatron (lower set of curves) and LHC (upper set).
        For each set, the solid (dashed) refer to NLO(LO) cross
        sections.
             The respective Born-level initial states are indicated
             in each panel. The value of the $R$-parity violating coupling
             (see eqn.(\protect\ref{strange_initial})) has been set to be
             0.01 and the CTEQ5 parametrisation has been used.
        }
\label{fig:ss}
\end{figure}

We now consider other possibilities for resonance
production. Restricting ourselves to sneutrinos
for the moment, note that a non-zero value for some of the
other $\lambda'$ couplings could lead to alternate
tree-level processes such as:
\be
\barr{rcl}
\lambda'_{i22} &:& s + \bar s \ra \tilde \nu_i \ ,\\
\lambda'_{i12} \: {\rm or}\  \lambda'_{i21} &:& d + \bar s \ra \tilde \nu_i \ .
\earr
        \label{strange_initial}
\ee
Although the chirality structure of the interaction term
is different for  the two cases in the second line, it is
of no consequence for either the LO cross-section
or the NLO corrections.
In Fig.\ref{fig:ss}, we plot these
cross-sections\footnote{Of course,
        similar production mechanisms involving a $b$-quark in the
        initial state is possible too. However, owing to the small
        flux for the $b$-quark, these are not of much interest
        phenomenologically.}
for both the Tevatron and the LHC. For an identical
value of the \rp\ violating coupling, the total cross-section is
much smaller than that in Fig.\ref{fig:prod}. This is only
to be expected as the strange-quark is a part of the sea and consequently
its flux is much smaller than that for the $d$-quark. Thus, for
the LHC, we have the relation
$\sigma(d \bar d) > \sigma(d \bar s) > \sigma(s \bar d) > \sigma(s \bar s)$,
while for the Tevatron, the second inequality is replaced by an equality.

\begin{figure}[htb]
\vspace*{0.1em}
\centerline{
\epsfxsize=14cm\epsfysize=6.0cm
                     \epsfbox{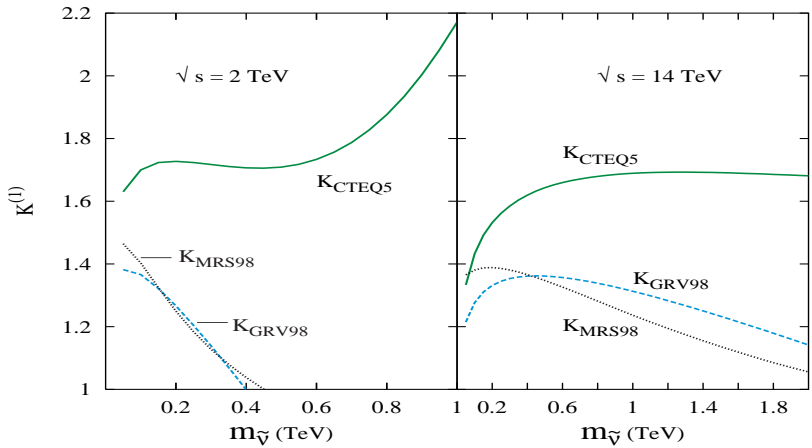}
}
\caption{\em The $K$ factor for the processes
            with coupling  $\lambda_{i 2 2}'$
                at the Tevatron
             as a function of the slepton (sneutrino) mass.
 }
\label{fig:K_fac_ss}
\end{figure}

As in the previous case, we may once again choose to parametrise
the NLO corrections in the form of a $K$-factor. And, although
we have chosen to present the cross-sections only for
the CTEQ5 parton distributions,
it is quite instructive to consider the dependence on the
parametrisation. In Fig.\ref{fig:K_fac_ss}, we present this for the
$s \bar s$ initial state.
The wide difference between the $K$-factor as calculated within
CTEQ5 \cite{cteq5},
with those obtained in the context of MRS98 \cite{mrs98} or
GRV98  \cite{grv98} may seem to be
a matter of concern. Interestingly,
unlike in the case of the $d \bar d$ initial state, the difference in $K$
here cannot be ascribed to the LO parton distributions. Rather, the
blame lies on the NLO parton distributions, in particular
the much larger strange-quark flux in the CTEQ5 parametrisation
(as compared to GRV98 or MRS98). Although this large deviation is partly
offset by a sizable negative contribution from the Compton diagram,
the latter effect is clearly subdominant. While such a discrepancy
might seem vexing, it is easier to appreciate once one considers the
experimental inputs in the parton density parametrisations, especially in the
large $x$ region. For example, the CTEQ collaboration uses
jet measurement data whereas MRS used prompt photon data.
And since our curves for the Tevatron reach unto a much larger
$x$ value than those for the LHC, the difference is more pronounced
in the former case. Notwithstanding this post-facto rationalisation,
the resultant  $K_{\rm CTEQ5}$ remains uncomfortably large, and moreover,
does not approach unity as $m_{\tilde \nu} \rightarrow \sqrt{s_{p \bar p}}$.
This is symptomatic of an inherent problem in the CTEQ5 parametrisation
for the heavier sea quark distributions.

\begin{figure}[htb]
\centerline{
\epsfxsize=14cm\epsfysize=6.0cm
                     \epsfbox{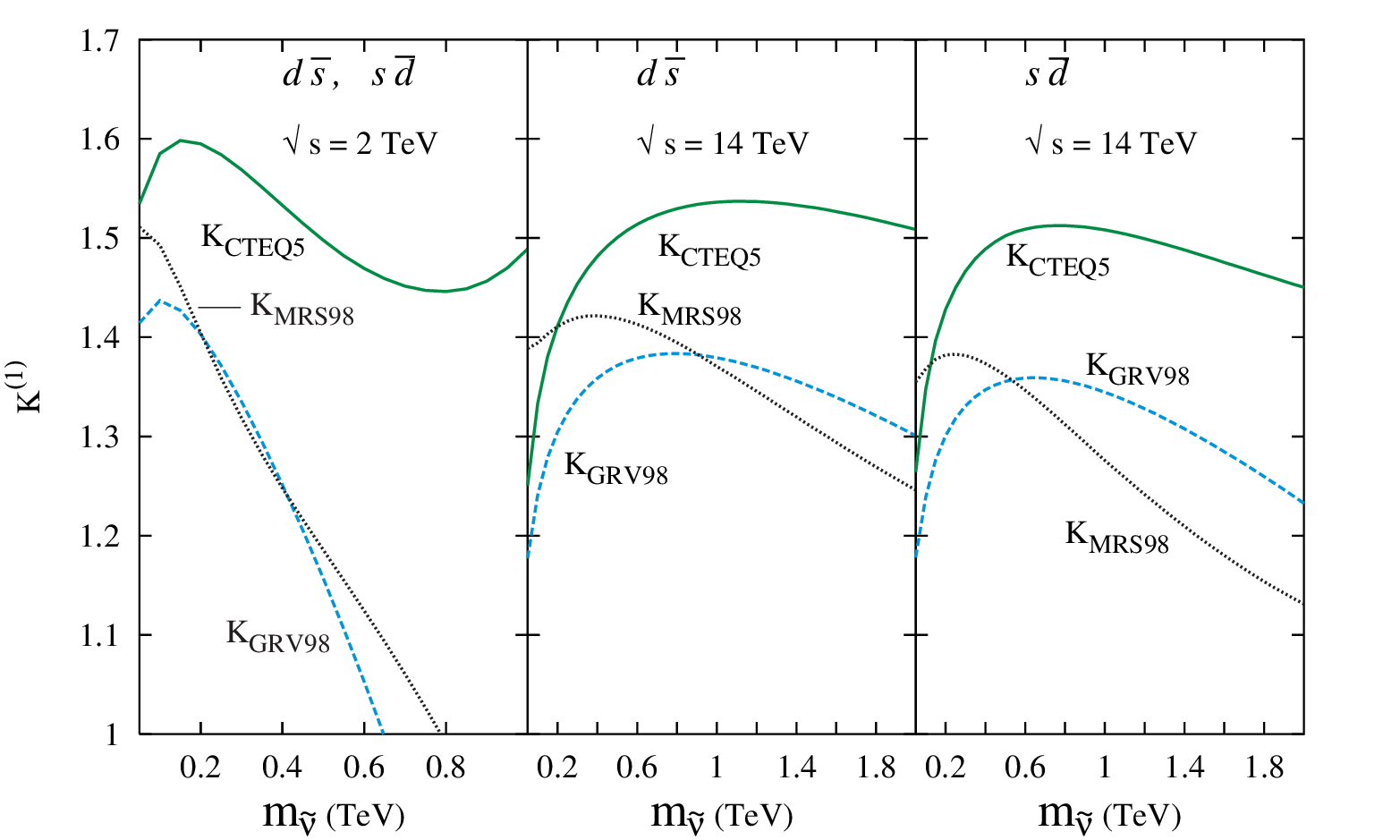}
}
\caption{\em The $K$ factor for the processes
        with coupling $\lambda_{i 1 2}'$ or $\lambda_{i 2 1}'$
                at the Tevatron
             as a function of the slepton (sneutrino) mass.
 }
\label{fig:K_fac_sd}
\end{figure}

The same problem is also reflected in the $K$-factors for the two
other cases of eqn.(\ref{strange_initial}) namely those with
$s \bar d$ and $d \bar s$ initial states. Once again
$K_{\rm MRS98}$ and $K_{\rm GRV98}$ are quite similar while
$K_{\rm CTEQ5}$ is significantly different (see Fig.\ref{fig:K_fac_sd}).

\subsection{Charged slepton production}
We finally consider slepton production. Governed by eqn.(\ref{lambda-pr}),
the relevant piece of the interaction Lagrangian is readily
seen to have the same structure as the piece responsible for
sneutrino production. Once again, various combinations of
quark-anti-quark pair could feature in the production process. However,
for brevity's sake we shall confine ourselves to a discussion of only
the processes corresponding initiated, at the Born level, by
$u \bar d$ (and $d \bar u$).  Of course, in the context of the
Tevatron, the two cross-sections would be identical,
whereas in the LHC, the former would be slightly larger.

\begin{figure}[!h]
\centerline{
\epsfxsize=17cm\epsfysize=7.0cm
                     \epsfbox{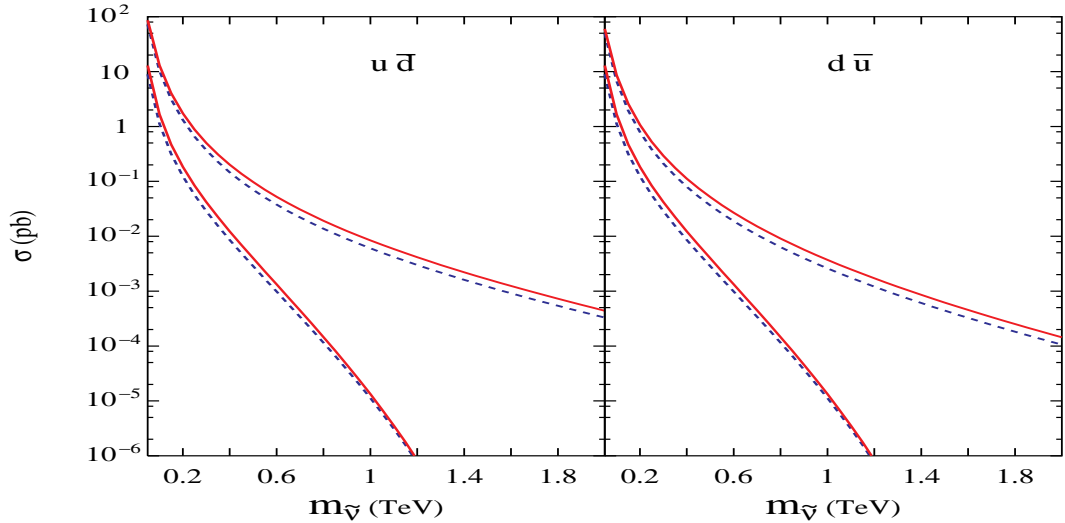}
}
\caption{\em Cross-section for charged slepton production at the
        Tevatron (lower set of curves) and LHC (upper set).
        For each set, the solid (dashed) refer to NLO(LO) cross
        sections.
        The respective Born-level processes are given by
             $ d + \bar{u} \rightarrow \tilde \ell^-$ and
             $ u + \bar{d} \rightarrow \tilde \ell^+$.
        The value of the $R$-parity violating coupling
             has been set to
             0.01 and the CTEQ5 parametrisation has been used.
 }
    \label{fig:prod_ubd}
\end{figure}

\begin{figure}[htb]
\centerline{
\epsfxsize=14cm\epsfysize=6.0cm
                     \epsfbox{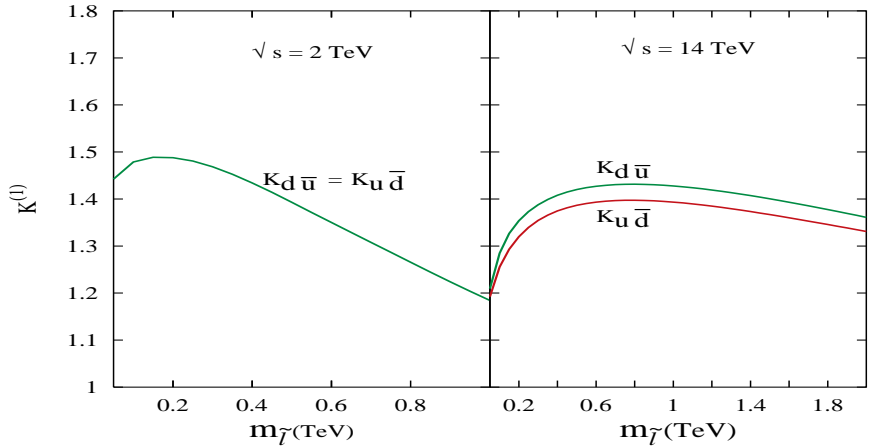}
}
\caption{\em The $K$ factor for charged slepton production
        (processes of Fig.\protect\ref{fig:prod_ubd})
        at the Tevatron as well as at the LHC
        as a function of the slepton mass. The CTEQ5 densities
        have been used.
 }
\label{fig:K_fac_ubd}
\end{figure}

As Fig.\ref{fig:prod_ubd} demonstrates, the behaviour of the cross-section
is identical to that of sneutrino production in $d \bar d$ collision (although
the magnitude is somewhat larger). This was only to be expected as
the flux of the $u$-quark inside the proton is similar to that
for the $d$-quark. To be very precise, the
valence $u$-density is a bit higher than the valence $d$, whereas the
sea-densities are very similar for the two.
Consequently, we should expect the
behaviour of the $K$-factor to be quite similar again, as is borne out by
Fig.\ref{fig:K_fac_ubd}. The remarkable similarity between
$K_{u \bar d}$ (and hence $K_{d \bar u}$ too for the Tevatron case)
and $K_{d \bar d}$ (Fig.\ref{fig:K_fac}) is a testimonial to the fact
that the only difference between the two cases arises from the
small effect due to isospin violation in the valence quark densities
(which, of course manifests itself primarily in the large scalar mass
region).

\section{Conclusions}
        \label{sec:conclusions}
To summarise, we have calculated the NLO corrections to
the resonant sneutrino and slepton production cross-sections (within
\rpv-MSSM) in the context of Tevatron and LHC. We find that for
processes controlled, at the Born level, by first-generation quarks,
the ensuing $K$-factor is a fairly sensitive function of the
scalar mass. For masses below about 1 TeV, the correction
can vary between 10\% and 50\% at the Tevatron with the correction
falling steeply for higher masses. At the LHC, the
mass-dependence is reduced significantly, and for masses less than
2 TeV, varies only between 1.2 and 1.45. While there is a significant
dependence on the structure function used, the effect is much
less pronounced for the NLO calculation than for the LO.
This lends us hope that once the next order effects are incorporated the
theoretical ambiguity would reduce to insignificant levels.

For production processes involving quarks of higher generations, the situation
is not so simple. The $K$-factors could be much larger, and worse, show a
marked dependence on the particular density parametrisation used. This is
but a reflection of the fact that these distributions are known with much
less precision and hence vary significantly between parametrisations.
Since the production cross-sections themselves are large enough
to be interesting, an NNLO calculation thus seems to be called for.

Finally, the calculations presented in this paper are not particular to
supersymmetric theories, but can be applied to any
color-singlet scalar (pseudoscalar) coupling to a quark-anti-quark pair.

\section*{Acknowledgements}
We would like to thank Anindya Datta for many useful discussions
and, more particularly, for his participation during the early
stages of the project. DC and VR would like to thank the Theory Division, 
CERN for hospitality while part of the project was being carried out.
DC also acknowledges the Department of Science and Technology, India for 
financial assistance under the Swarnajayanti Fellowship grant.

\newcommand{\plb}[3]{{Phys. Lett.} {\bf B#1} (#3) #2}                  %
\newcommand{\prl}[3]{Phys. Rev. Lett. {\bf #1} (#3) #2}        %
\newcommand{\rmp}[3]{Rev. Mod.  Phys. {\bf #1} (#3) #2}             %
\newcommand{\prep}[3]{Phys. Rep. {\bf #1} (#3) #2}                     %
\newcommand{\rpp}[3]{Rep. Prog. Phys. {\bf #1} (#3) #2}             %
\newcommand{\prd}[3]{{Phys. Rev.}{\bf D#1} (#3) #2}                    %
\newcommand{\np}[3]{Nucl. Phys. {\bf B#1} (#3) #2}                     %
\newcommand{\npbps}[3]{Nucl. Phys. B (Proc. Suppl.)
           {\bf #1} (#3) #2}                                           %
\newcommand{\sci}[3]{Science {\bf #1} (#3) #2}                 %
\newcommand{\zp}[3]{Z.~Phys. C{\bf#1} (#3) #2}                 %
\newcommand{\mpla}[3]{Mod. Phys. Lett. {\bf A#1} (#3) #2}             %
\newcommand{\astropp}[3]{Astropart. Phys. {\bf #1} (#3) #2}            %
\newcommand{\ib}[3]{{\em ibid.\/} {\bf #1} (#3) #2}                    %
\newcommand{\nat}[3]{Nature (London) {\bf #1} (#3) #2}         %
\newcommand{\nuovocim}[3]{Nuovo Cim. {\bf #1} (#3) #2}         %
\newcommand{\yadfiz}[4]{Yad. Fiz. {\bf #1} (#3) #2 [English            %
        transl.: Sov. J. Nucl.  Phys. {\bf #1} #3 (#4)]}               %
\newcommand{\philt}[3]{Phil. Trans. Roy. Soc. London A {\bf #1} #2
        (#3)}                                                          %
\newcommand{\hepph}[1]{(electronic archive:     hep--ph/#1)}           %
\newcommand{\hepex}[1]{(electronic archive:     hep--ex/#1)}           %
\newcommand{\astro}[1]{(electronic archive:     astro--ph/#1)}         %

\end{document}